\documentclass{article}
\pdfoutput=1

\usepackage{arxiv}
\usepackage{epstopdf}

\usepackage[utf8]{inputenc} 
\usepackage[T1]{fontenc}    
\usepackage{hyperref}       
\usepackage{url}            
\usepackage{booktabs}       
\usepackage{amsfonts}       
\usepackage{amsmath} 
\usepackage{nicefrac}       
\usepackage{microtype}      
\usepackage{textcomp}	    
\usepackage{graphicx}		
\usepackage{placeins}		
\usepackage{caption}		

\title{Modeling Electrical Resistance Drift with Ultrafast Saturation of OTS Selectors}

\author{
  Yi\u{g}it Demira\u{g}\\
  Department of Electrical and Electronics Engineering\\
  École Polytechnique Fédérale de Lausanne (EPFL)\\
  Lausanne, Switzerland \\
  \texttt{yigitdemirag@protonmail.ch} \\
  \And
  Ekmel \"{O}zbay \\
  Department of Electrical and Electronics Engineering\\
  Department of Physics \\
  Bilkent University\\
  Ankara, Turkey \\
  \texttt{ozbay@bilkent.edu.tr} \\   
  \And
 Yusuf Leblebici \\
  Department of Electrical and Electronics Engineering\\
  École Polytechnique Fédérale de Lausanne (EPFL)\\
  Lausanne, Switzerland \\
  \texttt{yusuf.leblebici@epfl.ch} \\
}

\begin{document}
\maketitle

\begin{abstract}
Crossbar array architecture is an essential design element for densely connected Non-Volatile Memory (NVM) applications. To overcome intrinsic sneak current problem of crossbar arrays, each memory unit is serially attached to a selector unit with highly nonlinear current-voltage (I-V) characteristics. Recently, Ovonic Threshold Switching (OTS) materials are preferred as selectors due to their fabrication compatibility with PRAM, MRAM or ReRAM technologies; however, OTS selectors suffer from the temporal drift of its threshold voltage. First, based on Poole-Frenkel conduction, we present time and temperature dependent model that predicts temporally evolving I-V characteristics, including threshold voltage of OTS selectors. Second, we report an ultrafast saturation (at $\sim10^3$ seconds) of the drift and extend the model to predict the time of drift saturation. Our model shows excellent agreement with OTS devices fabricated with 8 nm technology node at 25 \textdegree C and 85 \textdegree C ambient temperatures. The proposed model plays a significant role in understanding OTS device internals and the development of reliable threshold voltage jump table.
\end{abstract}

\keywords{OTS selectors \and Drift problem \and Device modeling \and Neuromorphic hardware}

\section{Introduction}
An OTS material is a thin-film two terminal amorphous chalcogenide, whom electrical conductivity can rapidly change from the high resistive state (HRS) to low resistive state (LRS) by applying large potential exceeding a specific threshold voltage ($V_{th}$). The conductivity difference between HRS and LRS can be as high as 4 orders of magnitude; nevertheless, the device immediately switches back when the applied potential is cut \cite{ovshinskyReversibleElectricalSwitching1968}. Its high ON/OFF current ratio and fast switching make OTS a promising candidate material for selector applications.

OTS is a chalcogenide material; hence it can crystallize. However, once crystallization starts, it is not feasible within device operation range to initiate melting and recover amorphous state. Therefore, crystallized OTS selector units are always assumed non-operational. As a solution, OTS materials can be carefully selected to have lower ionicity and higher hybridization, which lead to more directed covalent bonds to significantly slow crystallization process \cite{veleaTebasedChalcogenideMaterials2017}.

OTS selectors also perfectly match to physical and electrical scaling properties of NVM technologies. Owing to thin film compatibility with mature metallization techniques and CMOS stack support, densely connected 3D crossbar arrays with OTS selectors have been demonstrated \cite{derchangkauStackableCrossPoint2009}.

\section{The Drift Problem of OTS Selectors}
\label{sec:Head1}

\begin{figure}[h!]
  \centering
  \includegraphics[width=25pc]{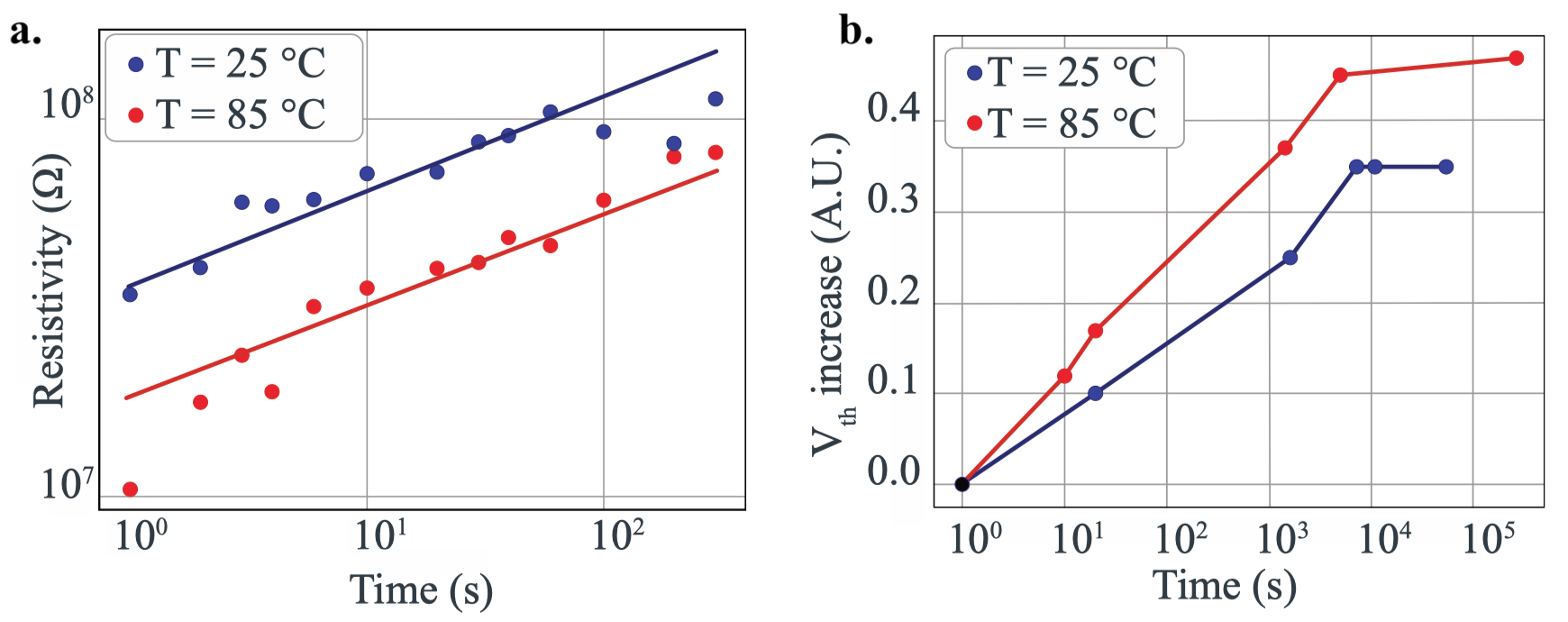}
  \caption{(a) Experimental (dot) and simulated (line) resistivity measurements of the same OTS device with 800 nm\textsuperscript{2} contact area at different ambient temperatures. The READ voltage values are 2.48 V and 2.64 V respectively for 25 \textdegree C and 85 \textdegree C experiments. The resistivity difference is due to thermally activated Poole-Frenkel behavior. (b) Resistance drift saturation measurements. The saturation of drift is faster at 85 \textdegree C due to faster annihilation of the defects.}
  \label{fig:fig1}
  \FloatBarrier
\end{figure}

The major problem of OTS technology is that the electrical conductivity of the selector decreases over time (Fig.~\ref{fig:fig1}(a)), called the drift problem. We observe that this conductivity decrease is not consistent, but saturates in time. Whether the conductance is drifting or already saturated, application of any potential higher than the threshold voltage ($V>V_{th}$), resets the drift and revert to selector's initial HRS level. Saturation of the drift is a rarely reported physical phenomenon in the literature, but critically important for understanding and developing OTS technology. In our OTS devices, we observe an ultrafast drift saturation (Fig.~\ref{fig:fig1}(b)) which takes places at least 2-3 orders of magnitude faster than reported drift measurements \cite{pirovanoElectronicSwitchingPhaseChange2004, legalloCompleteTimeTemperature2016}. 

The main problem is due to the increase of $V_{th}$ as conductance drifts. READ and WRITE operations require a known $V_{th}$ level of OTS. If $V_{th}$ increases and applied READ/WRITE pulses could not pass threshold value, then the selector device stays in HRS, hence READ/WRITE operations fail. One practical solution is to determine new $V_{th}$ with applying various prior READ voltages and detecting the threshold voltage value, which certain current level is reached \cite{sebastianMultilevelStoragePhasechange2016}. However, this solution requires additional support circuitry and increases the power consumption of the device; therefore, it is unfavorable. On the other hand, physical modeling of time-dependent resistance drift may lead to efficient solutions.

Modeling the drift behavior of OTS selectors is of vital importance for two reasons. 
First, a validated model can be useful for developing reliable time and temperature dependent jump-table of $V_{th}$. 
Second, a physically grounded model can provide a comprehensive understanding of temporally evolving non-measurable material properties such as activation energy $(E_a)$ and inter-trap distance $(\Delta z)$ to investigate the fabricated material in more detail.
Although there exist models capturing the drift behavior, these methods are either validated only for phase change memory (PCM) on a short range of time and ambient temperature or unable to predict the saturation of drift due to employing simple power-law-like models \cite{kimResistanceThresholdSwitching2011, ciocchiniModelingThresholdVoltageDrift2012} .

\begin{figure}[h!]
  \centering
  \includegraphics[width=17pc]{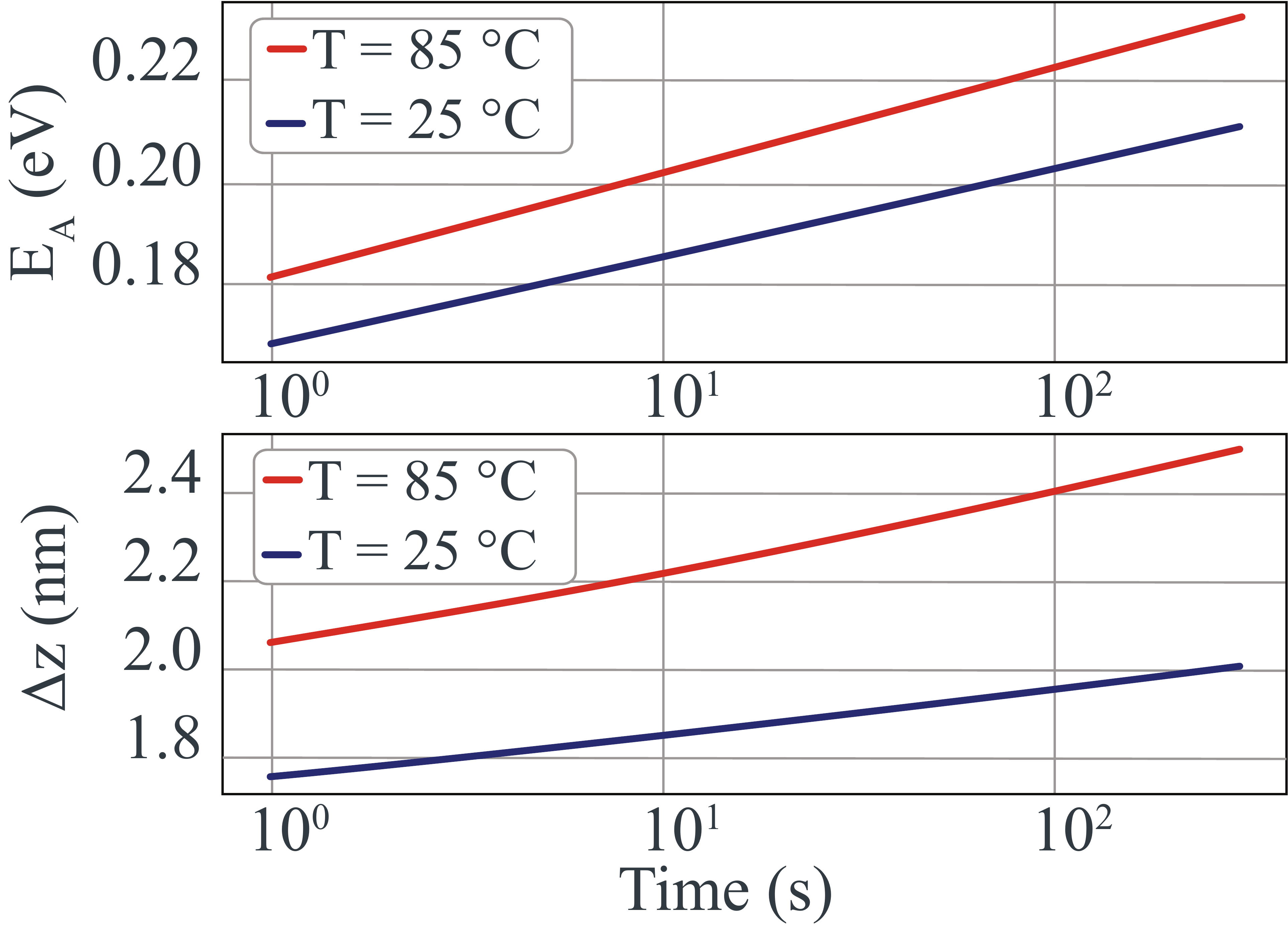}
  \caption{Temporally evolving activation energy $(E_a)$ and inter-trap distance $(\Delta z)$ calculated by the drift model. Upper and lower plots are generated for OTS selectors whom time and temperature dependent resistivity measurements are shown at Fig.~\ref{fig:fig1}}
  \label{fig:fig2}
  \FloatBarrier
\end{figure}

\newpage
\section{Modeling OTS Drift and Its Saturation}

\subsection{Time and Temperature Dependent Drift Model}

\begin{figure}[h!]
  \centering
  \includegraphics[width=35pc]{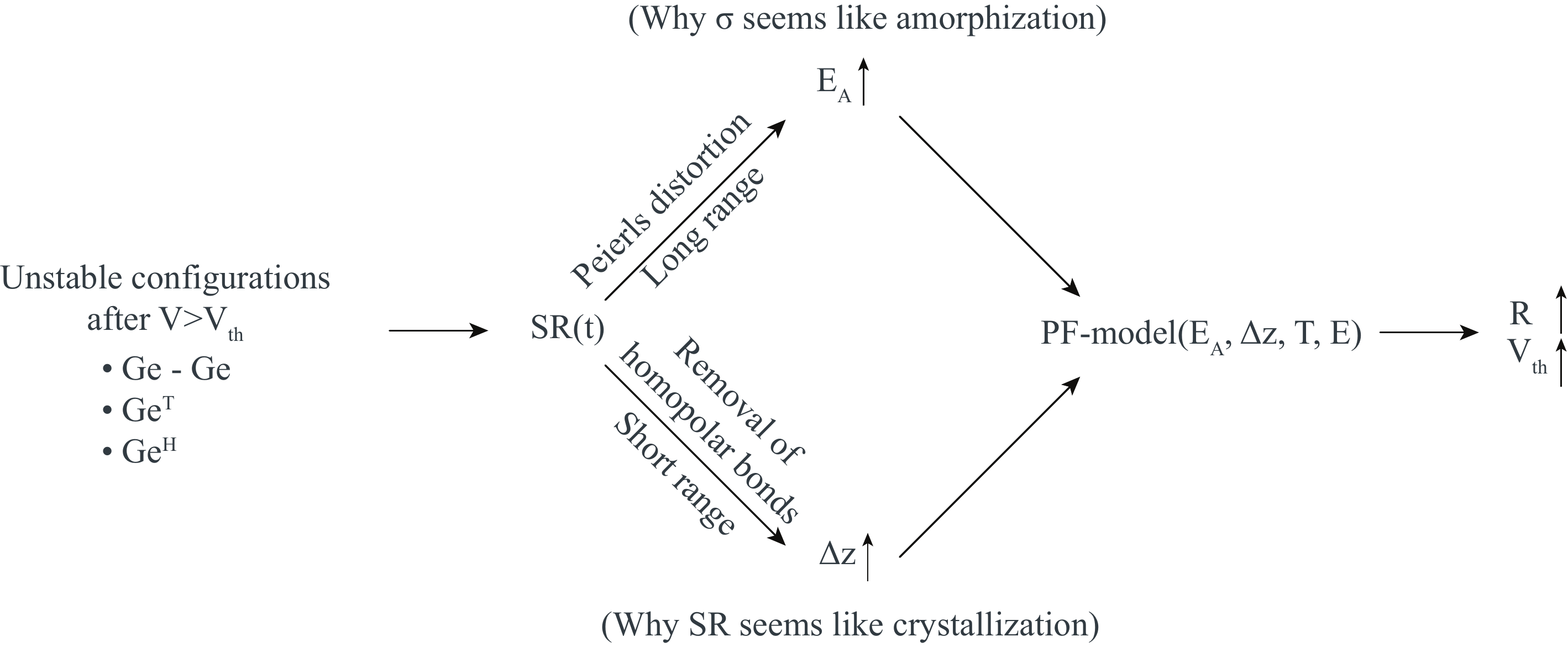}
  \caption{From left to right: Homopolar bonds in amorphous OTS are energetically unstable hence naturally disappear with time. This mechanism is called structural relaxation (SR) and affects activation energy ($E_a$) and inter-trap distance ($\Delta z$). Change of mainly these two causes the drift of device resistivity and threshold voltage as the electrical transport model of OTS materials follow the Poole-Frenkel conduction.}
  \label{fig:fig3}
  \FloatBarrier
\end{figure}

The physical phenomena governing the resistance drift on amorphous chalcogenide materials is yet to be fully understood. Raty et al. \cite{ratyAgingMechanismsAmorphous2015}, Gabardi et al. \cite{gabardiMicroscopicOriginResistance2015} and Zipoli et al. \cite{zipoliStructuralOriginResistance2016} have recently provided a significant insight into the microscopic picture of the the drift mechanism. Using \textit{ab-initio} simulations, it was found that there exist energetically unstable homopolar bonds and defects in melt-quenched amorphous. As these unstable defects naturally transform into lower-energy structures with time, the distance between intrinsic traps increase. The structure evolves into a more crystalline-like state, however without the necessary long-range order (Fig.~\ref{fig:fig3}).

To model the kinetics of the structural relaxation, we started with a drift model developed by Le Gallo et al., which has been previously validated on GeTe and GST \cite{legalloCompleteTimeTemperature2016}. In this model, the bond network state of chalcogenide is denoted with an order parameter $\Sigma$. $\Sigma$ is a normalized parameter between 0 (low-ordered highly stressed state) and 1 (ideal, energetically favorable relaxed state). Whenever $V>V_{th}$ is applied to OTS selector, amorphous network state resets and stressed with initial distance  $\Sigma(0)=\Sigma_0$ from the equilibrium. As network collectively relaxed through more favorable states with time, the energy barrier required to overcome, $E_b$, monotonically increases and it is assumed to be linearly dependent to $\Sigma$:

\begin{equation}
	E_{b}(t)=E_{s}(1-\Sigma(t)),
\end{equation}

where $E_{s}$ is the final energy barrier to reach the most relaxed state at $\Sigma=0$. With an Arrhenius type temperature-dependence, this relaxation occurs at the rate of $r(t)=v_{0} \exp \left(-E_{b}(t) / k_{B} T\right)$, where $v_0$ is an attempt-to-relax frequency, $k_B$ is the Boltzmann constant. After that the evolution of $E_b(t)$ can be calculated by:

\begin{equation}
	\frac{d \Sigma(t)}{d t}=-v_{0} \Delta_{\Sigma} \exp \left(-\frac{E_{b}}{k_{B} T(t)}\right).
\label{eq:re}
\end{equation}

At a constant temperature, Eq.~\ref{eq:re} can be solved analytically to track the progress of structural relaxation, such that

\begin{equation}
	\Sigma(t)=-\frac{k_{B} T}{E_{s}} \log \left(\frac{t+\tau_{0}}{\tau_{1}}\right),
\label{eq:SR}
\end{equation}

where $\tau_{1}=\left(k_{B} T / \nu_{0} \Delta_{\Sigma} \mathrm{E}_{\mathrm{s}}\right) \exp \left(E_{s} / k_{B} T\right)$ and $\tau_0=\tau_{1} \exp \left(-\Sigma_{0} E_{s} / k_{B} T\right)$. Once $\Sigma$ is calculated by Eq.~\ref{eq:SR}, an empirical linear relationship between structural relaxation, activation energy and inter-trap distance can be written as:

\begin{equation}
	\begin{array}{c}{E_{a 0}(t)=E^{*}-\alpha \Sigma(t),} \\ {\Delta z(t)=s_{0} / \Sigma(t),}\end{array}
\end{equation}

where $E^{*}$ is the activation energy at the equilibrium, $\alpha$ and $s_0$ are the constants linking change in $\Sigma$ to material properties. Finally, the temperature dependence of activation energy is assumed to follow the Varshni effect $\left(E_{a}=E_{a 0}-\xi T^{2}\right)$, as the optical bandgap of the material depends on the temperature \cite{varshniTEMPERATUREDEPENDENCEENERGY}. 

\subsection{Subthreshold Electrical Transport Model}
Amorphous chalcogenide materials are known to follow Poole-Frenkel subthreshold conductivity behavior \cite{hillHoppingConductionAmorphous1971}. The Poole-Frenkel effect suggests that thermal excitation and strong electric field release trapped carriers from ionizable defect centers, which are believed to create Coulomb potential. In this work, we used a previously developed field and temperature dependent 3D Poole-Frenkel emission model with field independent mobility \cite{galloSubthresholdElectricalTransport2015}. We first calculated the potential profile between defect centers located at $r=0$ and $r=\Delta z$ in all directions using Eq.~\ref{eq:PF_model}, where $\beta$ is the Poole-Frenkel constant, $e$ the electronic charge, $\theta$ the direction of escape relative to applied E-field $F$.

\begin{equation}
V(r, \theta, F)=-e F r \cos (\theta)-\frac{\beta^{2}}{4 e}\left(\frac{1}{r}+\frac{1}{\Delta z-r}\right)+\frac{\beta^{2}}{e \Delta z}
\label{eq:PF_model}	
\end{equation}

\begin{figure}[h!]
  \centering
  \includegraphics[width=19pc]{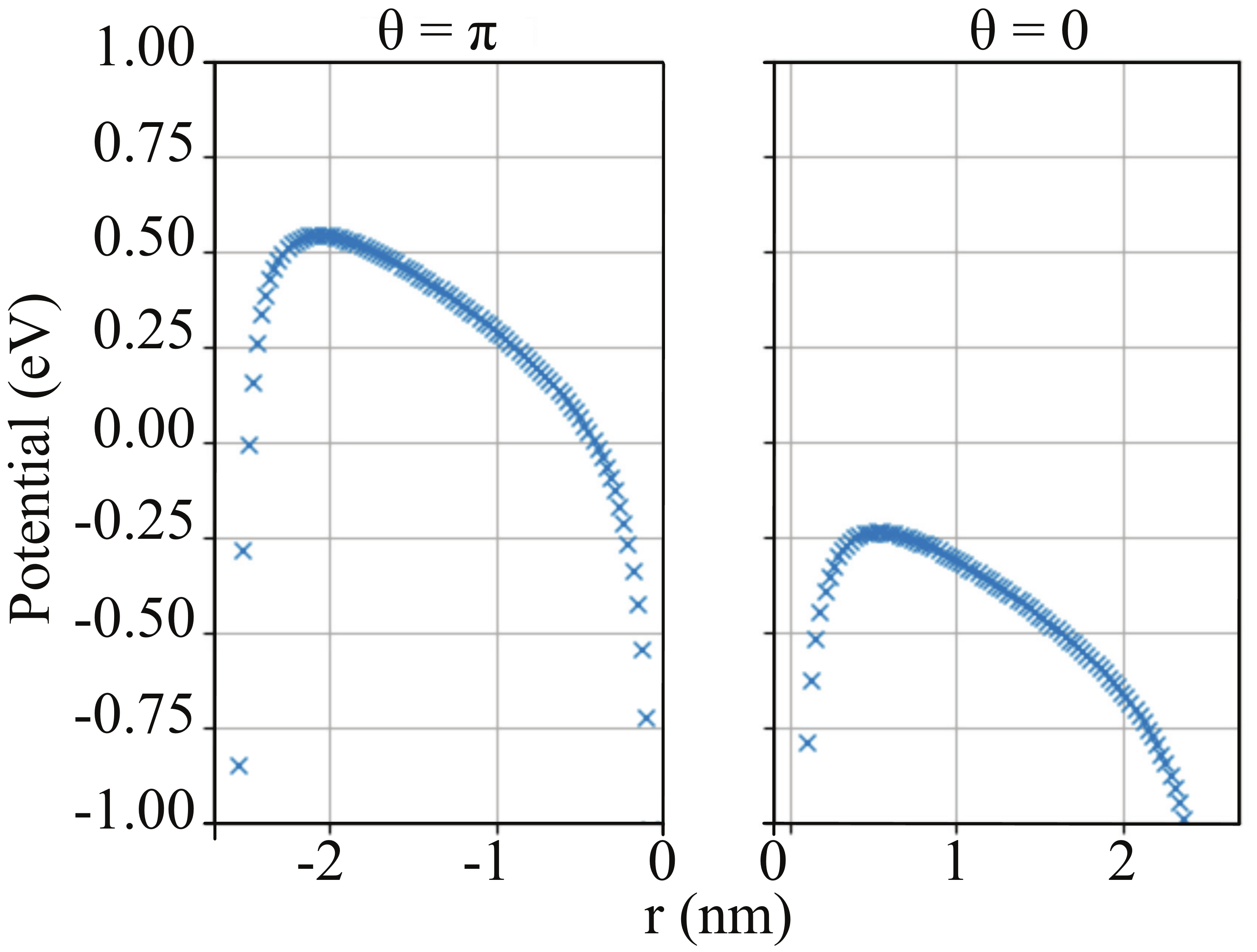}
  \caption{The potential profile between OTS defect centers are calculated via Eq.~\ref{eq:PF_model}. The amount of energy barrier may increase (left) or decrease (right) significantly due to the angle $(\theta)$ between strong electric field $3 \times 10^8$ V/m and the direction of escape.}
  \label{fig:figpf}
  \FloatBarrier
\end{figure}

The potential profile between Coulombic defect centers separated by $\Delta z$ is shown for OTS material in Fig.~\ref{fig:figpf}. The energy barrier lowering due to the Poole-Frenkel effect then can be calculated by

\begin{equation}
	E_{\mathrm{PF}}(F, \theta)=-\max _{r} V(r, \theta, F).
	\label{eq:barrierlowering}
\end{equation}

Finally, assuming Boltzmann statistics, we calculated the subthreshold electrical conductivity of the selector with:

\begin{equation}
	\sigma=e \mu \frac{K}{4 \pi} \int_{0}^{\pi} \exp \left(-\frac{E_{a}-E_{P F}(F, \theta)}{k_{B} T}\right) 2 \pi \sin (\theta) d \theta.
\label{eq:3D}
\end{equation}

\newpage
\subsection{Predicting Saturation Time of Electrical Resistance Drift}
In our experiments, we observed an unusually ultrafast saturation of the electrical resistance drift. Moreover, the drift saturation point changes as a function of the ambient temperature.

In the strong form of the drift model proposed by Le Gallo et al. \cite{legalloCollectiveStructuralRelaxation2018}, the evolution of subthreshold electrical resistance can be predicted; however, it falls short predicting a drift saturation point. To extend the previous model to predict the saturation time, we hypothesize that identical devices at different ambient temperatures that saturate at different times, eventually converge to the same $E_a$ and $\Delta z$ at the time of the saturation. This hypothesis requires $\Sigma(t)$ to be the same and constant for identical devices at different temperatures after the saturation time, $t_{SAT}$:

\begin{equation}
	\label{eq:saturation}
	\frac{d \Sigma(t)}{d t}=-r(t) \Delta_{\Sigma}=0 \hspace{1cm} \mathrm{for}~~t>t_{SAT}.
\end{equation}

\begin{figure}[h!]
  \centering
  \includegraphics[width=19pc]{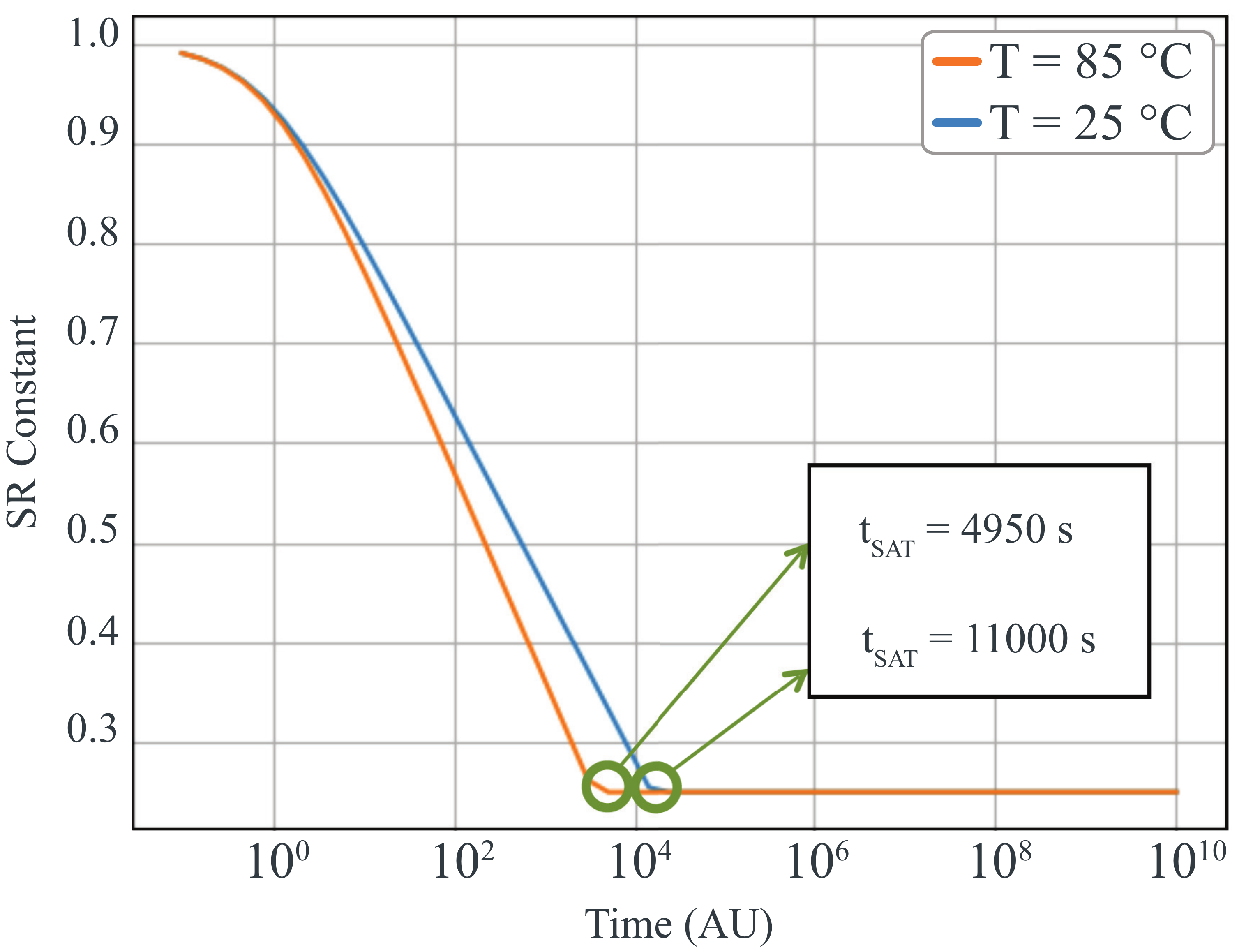}
  \caption{Calculated structural relaxation constant $\Sigma$ given for two ambient temperatures 25 \textdegree C and 85 \textdegree C.}
  \label{fig:SR_constant}
  \FloatBarrier
\end{figure}

During the training of the model, structural relaxation parameters are tuned according to this constraint given at Eq.~\ref{eq:saturation}. $t_{SAT}$ values used in the training are experimentally gathered as the saturation times of $V_{th}$ (see Fig.~\ref{fig:fig1} (b)). Figure~\ref{fig:SR_constant} shows the evolution of $\Sigma(t)$ for a trained model.

\subsection{Predicting Time Evolution of Threshold Voltage}
As the electrical resistance of the selector drifts with time the threshold voltage also drifts. To predict the time evolution of $V_{th}$, a mere time and temperature dependent subthreshold electrical model would not suffice. The model requires an extension to explain the moment of threshold switching for OTS.

\begin{figure}[h!]
  \centering
  \includegraphics[width=23pc]{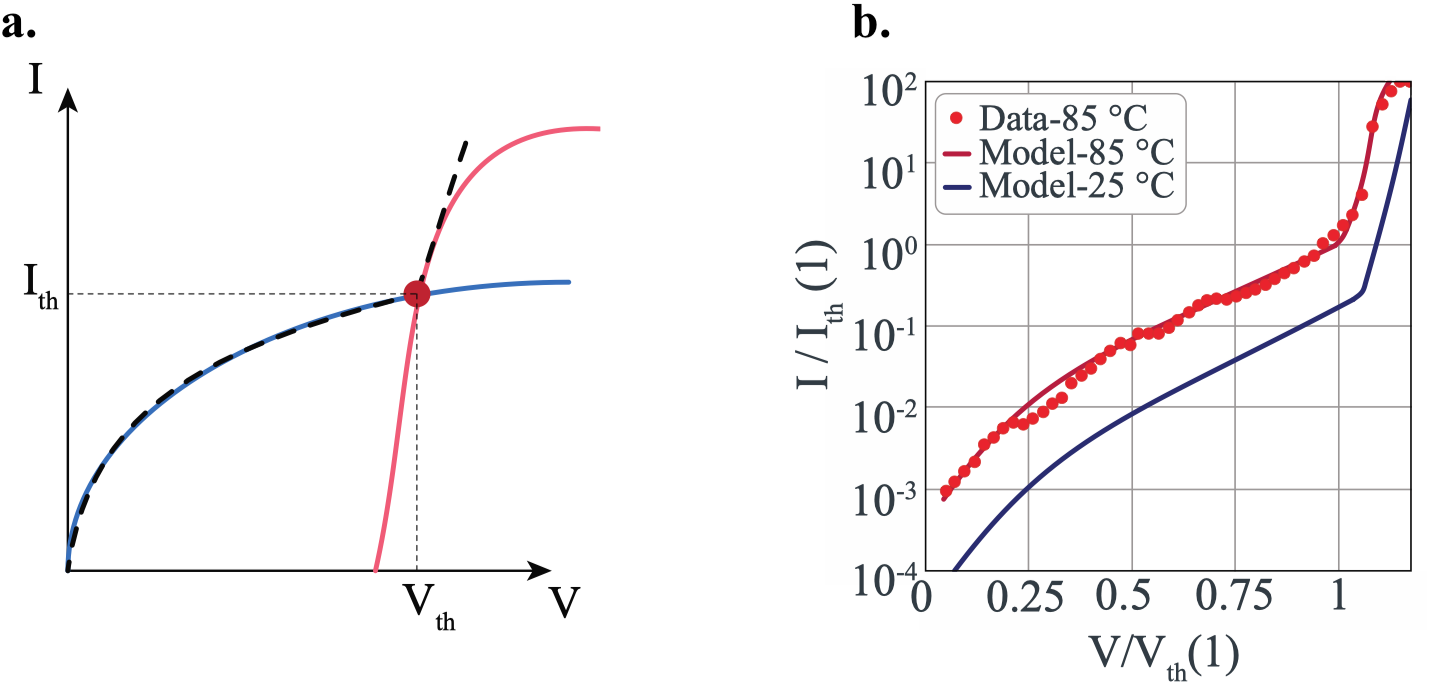}
  \caption{(a) Illustration of Poole-Frenkel (blue) conduction model and Okuto-Crowell impact ionization model (red). Together two models are sufficient for describing threshold switching of OTS. (b) Prediction of $V_{th}$ for T=25 \textdegree C, 85 \textdegree C and $t=300$ s.}
  
  \label{fig:ii}
  \FloatBarrier
\end{figure}

To account the sudden increase of conductivity during threshold switching, we combined Poole-Frenkel subthreshold transportation with Okuto-Crowell impact ionization. Okuto-Crowell impact ionization is an empirical model which is based on electron-avalanche multiplication effect due to the high electric field ($\sim 4\times 10^7$ V/m) inside the OTS material \cite{okutoThresholdEnergyEffect1975}. With this extension illustrated in Fig.~\ref{fig:ii} (a), we demonstrated a successful prediction of time-evolution of $V_{th}$ of OTS selectors at different ambient temperatures (Fig.~\ref{fig:ii} (b)).

\section{Physically-Realistic Parameter Optimization}
\begin{figure}[h!]
  \centering
  \includegraphics[width=25pc]{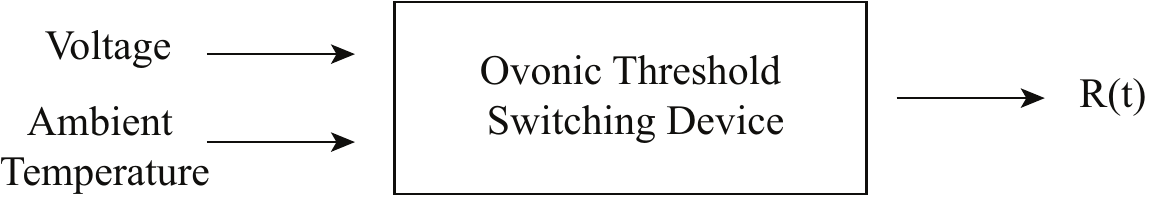}
  \caption{For OTS device, the only control parameters are ambient temperature and applied voltage.}
  \label{fig:rtheta}
  \FloatBarrier
\end{figure}

To adjust model parameters according to experimental measurements, the model I/O is matched with experimental conditions. The implemented model takes the same control inputs with the fabricated device (voltage, ambient temperature) and returns the same measurable quantity (resistivity). Figure~\ref{fig:rtheta} shows that fabricated OTS device can be modeled as a black box whom physical characteristics are represented by a set of parameters $\theta$.

\begin{figure}[h!]
  \centering
  \includegraphics[width=35pc]{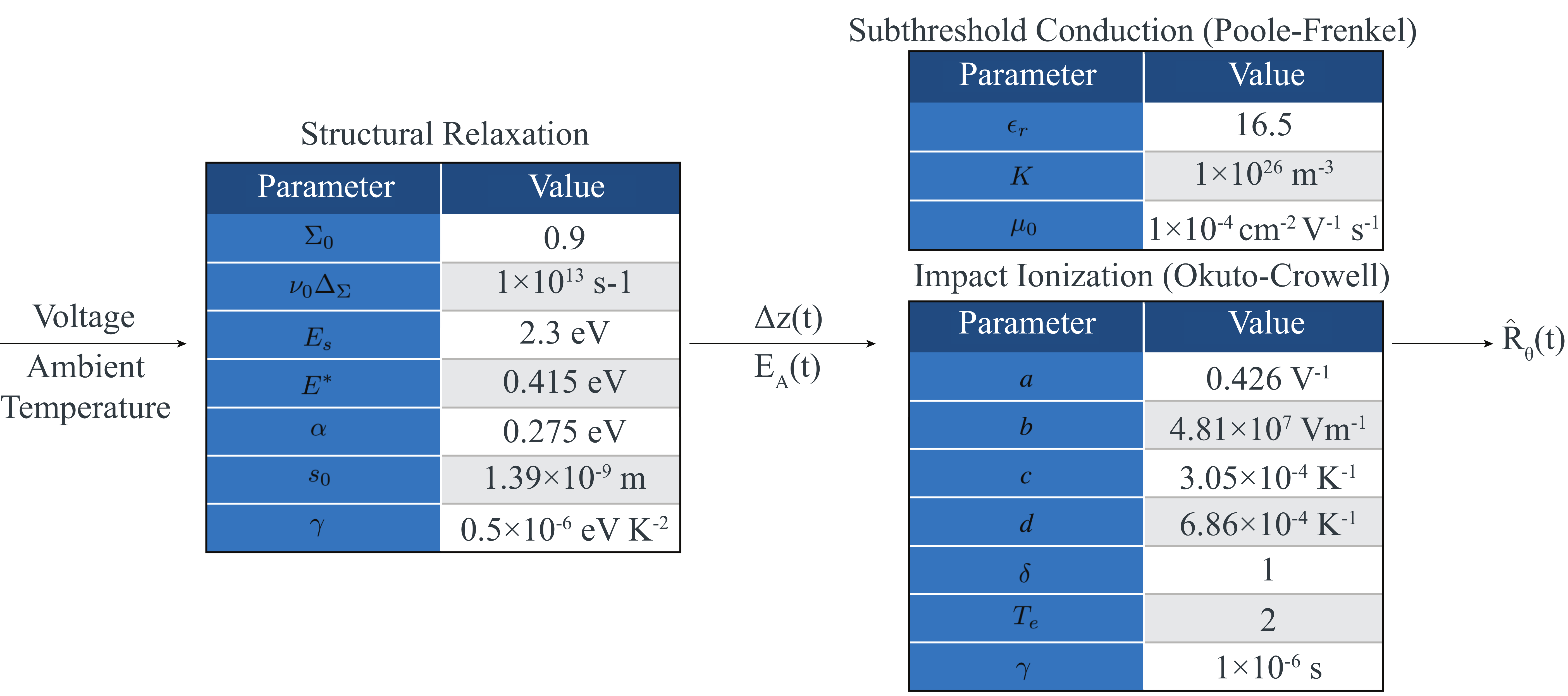}
  \caption{Model parameters used in this work. The temporal nature of electrical conductance is due to structural relaxation model, which affects $\Delta z$ and $E_A$ of OTS. Poole-Frenkel model and impact ionization combined, define the electrical behavior of the OTS and enable tracking of time-dependent $V_{th}$.}
  \label{fig:parameters}
  \FloatBarrier
\end{figure}

Figure~\ref{fig:parameters} shows the proposed drift saturation model with 17 parameters. To successfully optimize these model parameters to match the fabricated device, we consider two requirements. First, in an ideal situation, the proposed model and the fabricated device must output the same resistivity level when applied the same voltage and ambient temperature. Therefore, the aim is to minimize the difference between measured resistivity of the fabricated device, $R(\theta)$, and the modeled resistivity, $R(\hat{\theta})$, by tuning model parameters, $\theta$.

Second, the model parameters with physical correspondence must stay within their physically-realistic ranges. To limit every parameter with different upper (UB) and lower bounds (LB) as in Eq.~\ref{eq:e3}, several optimization methods could be used e.g., simulated annealing, evolutionary or gradient-based search algorithms. We utilized simulated annealing for its easy implementation and despite its computationally-heavy search, it successfully minimized the loss function inside of Eq.~\ref{eq:e3} with physically-realistic parameters \cite{metropolisEquationStateCalculations1953}.

\begin{align}
	x = \arg \min _ { \theta } \frac { 1} { 2} \sum \left( R ( t ) - \hat { R } _ { \theta } ( t ) \right) ^ { 2} ,\text{ subject to } L B _ { \text{i} } < \theta _ { i } < U B _ { i }.
	\label{eq:e3}
\end{align}

\section{Conclusion}
We reported an ultrafast saturation phenomenon (at $\sim10^3$ seconds) of resistance drift on OTS materials, which are promising selector candidates in the next generation NVM (PRAM, MRAM and ReRAM) crossbar technologies. An electrical transport model is proposed to describe time and temperature dependent OTS I-V characteristics. The model based on structural relaxation, Poole-Frenkel conduction, and impact ionization, is shown to be in close agreement with our devices fabricated with 8 nm node technology and tested at 25 \textdegree C and 85 \textdegree C ambient temperatures for $\sim 10^4$ seconds. The models and physical parameters (including $E_{a}$ and $\Delta z$) provide valuable insight into non-measurable material properties. With the support of drift saturation and $V_{th}$ prediction, our model may play a significant role in the development of reliable $V_{th}$ jump tables.

\newpage
\bibliographystyle{unsrt}  
\bibliography{references.bib}  
\end{document}